\documentclass[prb,preprint,showpacs,superscriptaddress]{revtex4}
\usepackage{graphicx}
\begin{document}
\title{Critical currents in weakly textured MgB$_2$: Nonlinear transport in anisotropic heterogeneous media.}

\author{M.~Eisterer}
\address{Atominstitut, Vienna University of Technology, Austria}
\author{W.~H\"a\ss ler}
\address{Institute for Metallic Materials, IFW Dresden, Germany}
\author{P.~Kov\'{a}\u{c}}
\address{Institute of Electrical Engineering, Slovak Academy of Sciences, Bratislava, Slovakia}

\begin{abstract}
A model for highly non-linear transport in heterogeneous media
consisting of anisotropic particles with a preferred orientation
is proposed and applied to the current transport in weakly
textured magnesium diboride, MgB$_2$. It essentially explains why,
unlike in conventional superconductors, a significant macroscopic
anisotropy of the critical currents can be induced by the
preparation of MgB$_2$ tapes. The field and angular dependence of
the critical current is calculated for various degrees of texture
and compared to experimental data.

\end{abstract}

\pacs{74.25.Sv,74.81.Bd,74.70.Ad,81.40.Ef}

\maketitle

\newpage

\section{Introduction}

The critical current in polycrystalline MgB$_2$ is a particular
example of nonlinear transport in heterogeneous media\cite{Sah98}.
The intrinsic upper critical field anisotropy results in different
critical current densities in differently oriented grains, when a
magnetic field is applied \cite{Eis03}. The problem was solved by
integration of the (local) current densities over the percolation
cross section, for which the distribution function of the local
critical currents has to be known. Although the local currents can
be described by standard pinning models \cite{Dew74}, the
resulting macroscopic currents are significantly altered. The
field dependence of the critical current increases and the current
becomes zero well below the upper critical field \cite{Eis03}.

In the present letter, a heterogeneous and macroscopically
anisotropic system is considered, i.e. the distribution function
of the relevant transport property (e.g., the critical current
density $J_\mathrm{c}$) depends on its orientation. This is
realized in weakly (or partially) textured MgB$_2$ and explains
the $J_\mathrm{c}$-anisotropy
\cite{Gra02,Son02,Kit05B,Kov05,Lez05b,Bei06,Lez06b,Kov08,Ser08},
which can exceed the \emph{intrinsic} anisotropy of MgB$_2$ (5--6)
\cite{Eis07rev} by orders of magnitude. The
$J_\mathrm{c}$-anisotropy can be induced by some preparation
techniques of MgB$_2$ tapes, which was impossible in the
conventional superconductors NbTi and Nb$_3$Sn since they are
intrinsically isotropic. On the other hand, high temperature
superconductors must be highly textured for enabling large
transport currents. Thus, MgB$_2$ is the first superconductor
needing a model for partial texture as description. The
distribution function of $J_\mathrm{c}$ in MgB$_2$ will be derived
in the following, but it is straightforward to apply the model to
any other highly nonlinear transport phenomena in anisotropic
heterogeneous media.

\section{Experimental\label{secexp}}

Powders of magnesium (Goodfellow, 99.8\%) and amorphous boron
(Starck, 95--97\%) were milled together for 50 hours  at 250\,rpm
in a planetary ball mill (Retsch PM 4000) under protecting argon
atmosphere. The powder was filled into iron tubes and deformed by
swaging and rolling into tapes and heat treated. One tape was
doped with nanostructured carbon (99.999\%), which had been added
to the precursor powders before milling (nominal
MgB$_{1.87}$C$_{0.13}$).

The critical current was measured by direct transport measurements
in liquid helium. The texture was characterized by measuring
rocking curves of the MgB$_2$(002) reflection using high energy
synchrotron X-ray ($E=77.5$\,keV) diffraction. Details of all
measurements and the sample preparation are given elsewhere
\cite{Hae09}, since we will focus on the proposed model, while the
experimental data are used for comparison.

\section{Mathematical Model\label{SecModel}}

\begin{figure} \centering \includegraphics[clip,width=0.4\textwidth]{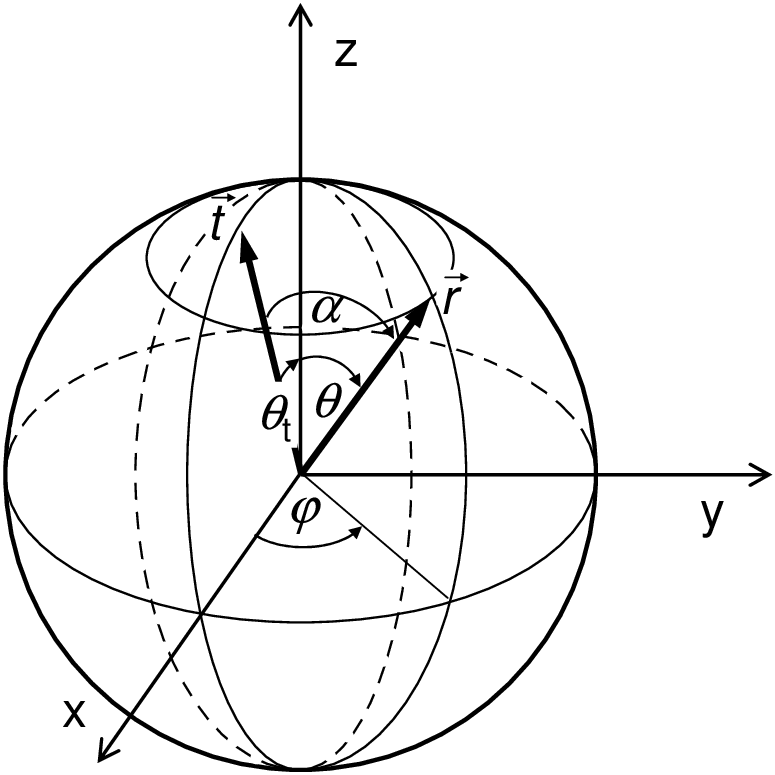}
\caption{Spherical coordinate system. The magnetic field is
assumed to be parallel to the z-axis. $\theta$ and $\varphi$
denote the polar and azimuthal angle, respectively. The preferred
orientation (texture) of the grains lies in their c-axis parallel
to $\vec{t}$.} \label{Figcoord}
\end{figure}

The macroscopic critical current density in anisotropic
polycrystalline superconductors (or any other transport property
in heterogeneous media, which hardly depends on the driving force)
can be calculated from \cite{Eis03}
\begin{equation}
J_\mathrm{c}^\mathrm{macro}=\int_0^{J_\mathrm{max}}(\frac{p(J)-p_\mathrm{c}}{1-p_\mathrm{c}})^{1.76}dJ
\label{Jcint}.
\end{equation}
The percolation threshold $p_\mathrm{c}$ is defined as the minimum
fraction of superconducting grains for a continuous current path.
$p(J)$ is the fraction of grains with a (local) $J_\mathrm{c}$
exceeding $J$ and $J_\mathrm{max}$ is given by the condition
$p(J_\mathrm{max})=p_\mathrm{c}$. The field dependence of the
(local) critical currents is assumed to follow the well
established model for grain boundary pinning~\cite{Dew74}
\begin{equation}
J_\mathrm{c}^\mathrm{local}(B)\propto \frac
{(1-\frac{B}{B_\mathrm{c2}})^2} {\sqrt{B}} \label{Jcmodel}
\end{equation}
for $B<B_\mathrm{c2}$ and $J_\mathrm{c}^\mathrm{local}=0$
otherwise. For the sake of simplicity, any difference between the
upper critical field, $B_\mathrm{c2}$, and the irreversibility
field, $B_\mathrm{irr}$, is neglected. The angular dependence of
the critical current is modeled by the scaling approach proposed
by Blatter et al.~\cite{Bla92}, i.e. $B$ is multiplied by
$\sqrt{\cos^2\theta+\gamma^{-2}\sin^2\theta}$, where $\gamma$
denotes the anisotropy of the upper critical field
$B_\mathrm{c2}^{ab}/B_\mathrm{c2}^c$ and $\theta$ the angle
between the applied magnetic field and the crystallographic
$c$-axis; thus, $B_\mathrm{c2}$ in Eq.~(\ref{Jcmodel}) corresponds
to $B_\mathrm{c2}^{c}$, the upper critical field for $B\|c$.

The angular distribution of the grain orientation,
$p_\mathrm{a}(\theta)$, is needed for calculating $p(J)$. Each
grain is represented by a point on the surface of the unit sphere
($\vec{r}$ in Fig.~\ref{Figcoord}), which corresponds to the
orientation of its $c$-axis. The density of these points,
$f_\mathrm{a}(\theta,\varphi)$ is per definition constant for a
random grain orientation. The number of grains with a polar angle
equal to or above $\theta$ is obtained from integration of
$f_\mathrm{a}(\theta,\varphi)$ over the corresponding surface area
\begin{equation}
p_\mathrm{a}(\theta)=\int_\theta^{\frac{\pi}{2}}\int_0^{2\pi}
f_\mathrm{a}(\theta',\varphi)d\varphi\ \sin{\theta'} d\theta'
\label{intfa},
\end{equation}
which reduces to $2\pi f_\mathrm{a}\cos{\theta}$ for constant
$f_\mathrm{a}$. The sine of $\theta'$ is the radius of the circle,
along which the integration over $\varphi$ is carried out (cf.
Fig.~\ref{Figcoord}). Due to the symmetry of the problem, the
integration can be restricted to the upper half of the unit
sphere. The relative fraction of such oriented grains is obtained
by setting $f_\mathrm{a}=1/2\pi$ (the surface of the unit sphere
is 4$\pi$). This leads to $p_\mathrm{a}(\theta)=\cos{\theta}$ and
to the correct result $p_\mathrm{a}(0)=1$ for a random grain
orientation.

Partial texture can be introduced by a varying density
$f_\mathrm{a}(\theta,\varphi)$. For uniaxial texture, the density
becomes independent of the azimuthal angle ($\varphi$), if the
preferred orientation of the grains' $c$-axis is assumed to be
parallel to the z-axis. A natural choice for the variation along
the polar angle is a Gaussian (or normal) distribution
\begin{equation}
f_\mathrm{a}(\theta,\varphi) \propto
\exp(-\frac{\theta^2}{2\alpha_\mathrm{t}^2}) \label{Gauss}.
\end{equation}
with standard deviation $\alpha_\mathrm{t}$. The proper
normalization is obtained as above by integration of
$f_\mathrm{a}(\theta,\varphi)$ over half the unit sphere and
$p_\mathrm{a}(\theta)$ can then be calculated from
Eq.~(\ref{intfa}). The chosen density function was confirmed by
synchrotron x-ray diffraction with $\alpha_\mathrm{t}=29.3^\circ$
\cite{Hae09}, but it is straightforward to use any other
distribution.

If the direction of the applied magnetic field is chosen as the
z-axis, the preferred orientation of the grains' $c$-axis is in
general not parallel to this direction. The situation is
illustrated in Fig.~\ref{Figcoord}. The preferred orientation is
defined by $\vec{t}$ with spherical coordinates
($\theta_\mathrm{t},\varphi=0$). The angle, $\alpha$, between
$\vec{t}$ and an arbitrary direction $\vec{r}$ is given by
\begin{equation}
\cos\alpha=\mathrm{abs}(\sin\theta\cos\varphi\sin\theta_\mathrm{t}+\cos\theta\cos\theta_\mathrm{t})
\label{alpha}.
\end{equation}
The density function $f_\mathrm{a}(\theta,\varphi)$ is obtained
from Eq.(\ref{Gauss}) by replacing $\theta$ by $\alpha$. It
depends on the polar and azimuthal angle and the two dimensional
integration~(\ref{intfa}) has to be calculated numerically.

At fixed magnetic field, $p(J)$ can be calculated from
$p_\mathrm{a}(\theta)$, since $J_\mathrm{c}^\mathrm{local}$
depends monotonically on $\theta$. Finally,
$J_\mathrm{c}^\mathrm{macro}$ is obtained by numerical evaluation
of integral~(\ref{Jcint}).

\section{Results and Discussion}

\begin{figure} \centering \includegraphics[clip,width=0.5\textwidth]{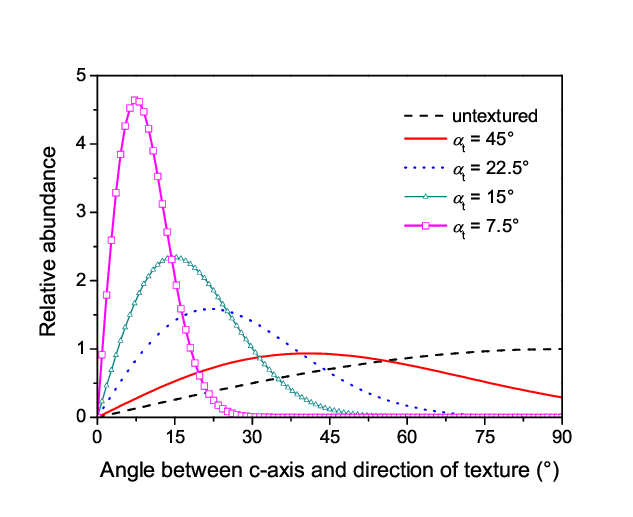}
\caption{(Color online) Grain orientation with respect to the
direction of partial texture. The maximum shifts continuously from
the perpendicular orientation (90$^\circ$) for random grain
orientation to zero in the limit of perfect texture
($\alpha_\mathrm{t}\rightarrow 0$).} \label{Figrelab}
\end{figure}

The relative abundance of grains, whose $c$-axis is tilted by an
angle $\theta$ from the direction of the texture $\vec{t}$, is
obtained by integration of $f_\mathrm{a}(\theta,\varphi)$ over
$\varphi$, which is the negative derivative of
$p_\mathrm{a}(\theta)$. It is plotted in Fig.~\ref{Figrelab} for
various angles $\alpha_\mathrm{t}$, which characterize the degree
of texture. With increasing texture, the maximum shifts from
$\pi/2$ for randomly oriented grains to smaller angles. (In the
limit $\alpha_\mathrm{t}\rightarrow 0$ (perfect texture), the
distribution function becomes a $\delta$-function.) For weak
texture, the majority of grains is \emph{not} oriented with the
c-axis close to the direction of the uniaxial texture, but the
fraction of grains with perpendicular orientation is reduced and
becomes negligible for $\alpha_\mathrm{t}\approx 25^\circ$.

\begin{figure} \centering
\includegraphics[clip,width=0.5\textwidth]{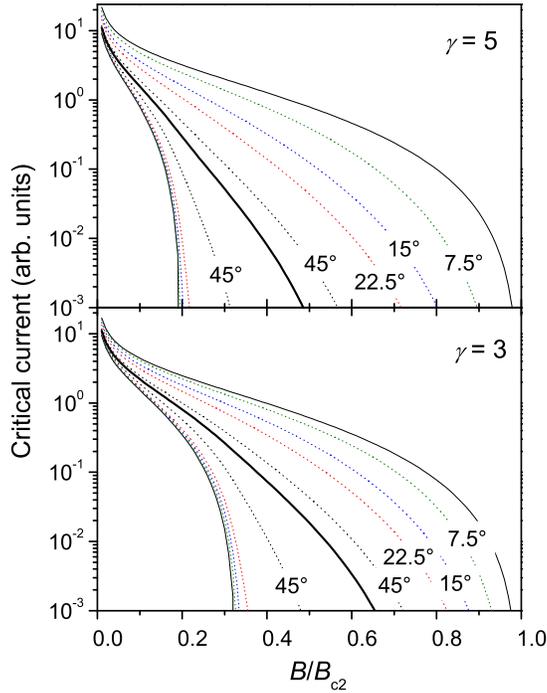}
\caption{(Colour online) Influence of partial texture on the
critical currents. The given angles refer to $\alpha_\mathrm{t}$.
The central bold lines represent untextured MgB$_2$. The dotted
lines below and above the bold lines correspond to the field
orientation parallel and perpendicular to the direction of
texture. The limiting curves represent perfect texture.}
\label{FigJccalc}
\end{figure}

The influence of partial texture on the field dependence of the
critical currents is shown in Fig.~\ref{FigJccalc}. The two panels
refer to different anisotropies; 5 is typical for clean MgB$_2$, 3
for moderately dirty MgB$_2$. \cite{Eis07rev} The magnetic field
is normalized by $B_\mathrm{c2}^{ab}$. $B_\mathrm{c2}^c$ is
$1/\gamma$ (0.2 and $0.\dot{3}$). The percolation threshold was
fixed to 0.25, which is an average value in polycrystalline
MgB$_2$ \cite{Eis03,Bei06,Yam07,Eis09}. The central bold line was
calculated for the untextured material, the graphs below and above
this line refer to the parallel ($\theta_\mathrm{t}=0^\circ$) and
perpendicular ($\theta_\mathrm{t}=90^\circ$) field orientation.
With decreasing $\alpha_\mathrm{t}$ (increasing texture) the
critical current becomes more anisotropic and the data converge
much faster to the limiting case of perfect texture (thin solid
lines) for the parallel orientation, where the difference between
perfect and partial texture becomes small at
$\alpha_\mathrm{t}\sim 20^\circ$. The $J_\mathrm{c}$-anisotropy
decreases with decreasing $\gamma$ but the qualitative behavior is
the same in both panels.

\begin{figure} \centering
\includegraphics[clip,width=0.5\textwidth]{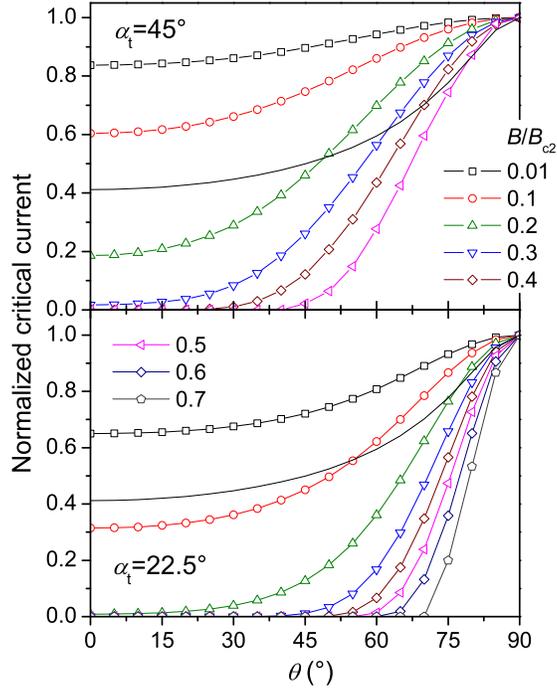}
\caption{(Color online) Angular dependence of the critical
current. $\gamma=5$ and $p_\mathrm{c}=0.25$ are assumed. The
texture is stronger in the lower panel. The graphs without symbols
refer to perfect texture (anisotropic scaling approach
\cite{Bla92}) at $B/B_\mathrm{c2}=0.01$. The critical current
anisotropy increases with field.} \label{FigJcang}
\end{figure}

The angular dependence of the critical currents is plotted in
Fig.~\ref{FigJcang}. The assumed texture is stronger in the lower
panel, the anisotropy ($\gamma=5$) and $p_\mathrm{c}$ (0.25) were
kept constant. At low magnetic fields, the
$J_\mathrm{c}$-anisotropy is small for the more weakly textured
material (1.2 at $B/B_\mathrm{c2}=0.01$), but increases with
texture (to $\sqrt{\gamma}$ for $B\rightarrow 0$ for perfect
texture, line graph in Fig.~\ref{FigJcang}). The
$J_\mathrm{c}$-anisotropy monotonically increases with magnetic
field until it diverges for any $\alpha_\mathrm{t}<\infty$, when
the currents reduce to zero in the parallel orientation
($\theta_\mathrm{t}=0^\circ$). This was observed experimentally
\cite{Kov05,Lez06b,Kov08}. (Note that $\vec{t}$ is perpendicular
to the surface of MgB$_2$ tapes, hence the parallel orientation
($\theta_\mathrm{t}=0^\circ$) corresponds to the field
perpendicular to the tape.) At fixed $\gamma$, $p_\mathrm{c}$, and
$\alpha_\mathrm{t}$, the $J_\mathrm{c}$-anisotropy is only a
function of the reduced magnetic field $b=B/B_\mathrm{c2}$, which
explains its temperature dependence at fixed magnetic field: $b$
increases with temperature, since $B_\mathrm{c2}$ decreases.

\begin{figure} \centering
\includegraphics[clip,width=0.5\textwidth]{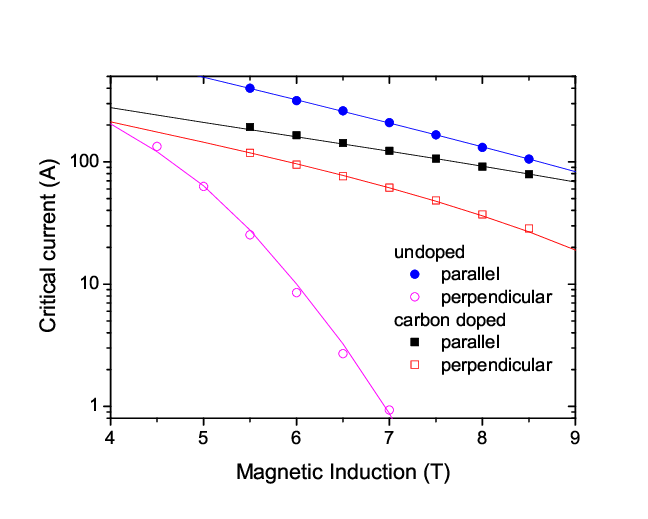}
\caption{(Color online) The model (line graphs) successfully
describes the measured critical currents (symbols) for both main
field orientations.} \label{FigJcexp}
\end{figure}

\begin{figure} \centering
\includegraphics[clip,width=0.5\textwidth]{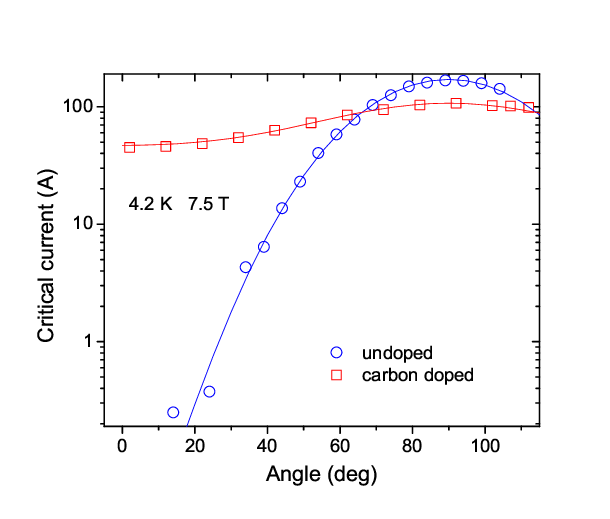}
\caption{(Color online) Calculated (lines) and experimental
(symbols) angular dependence of the critical currents at 7.5 T in
liquid helium.} \label{FigJcangexp}
\end{figure}

Experimental data (symbols) for the critical current are compared
to the prediction of the present model (line graphs) in
Figs.~\ref{FigJcexp} and~\ref{FigJcangexp}. The agreement is
excellent, given the numerous assumptions/simplifications (pure
grain boundary pinning \cite{Dew74}, anisotropic scaling
\cite{Bla92}, Gaussian distribution etc.). The correct choice of
the (unknown) parameters is difficult, since a 5 parameter fit
($B_\mathrm{c2}$, $\gamma$, $\alpha_\mathrm{t}$, $p_\mathrm{c}$,
and a constant prefactor, $I_\mathrm{c0}$) does not lead to
significant results. Therefore, $p_\mathrm{c}$ was fixed to 0.25
and $\alpha_\mathrm{t}$ to 29.3$^\circ$ \cite{Hae09}. The
remaining parameters were adapted to fit the experimental
$J_\mathrm{c}(B)$ data for both field orientations simultaneously.
$B_\mathrm{c2}=23.2 \, (20.5)$\,T, $\gamma=4.05 \, (1.8)$ and
$I_\mathrm{c0}=2840 \, (860)\,$A were chosen for the undoped
(doped) tape. The parameters of the doped tape
($T_\mathrm{c}=29.5$\,K) agree fairly well with results on a
carbon doped (9.5\%) single crystal with a similar
$T_\mathrm{c}=30.6$\,K \cite{Kru06}, where $\gamma$ slightly below
2 and $B_\mathrm{c2}(0)=25\,$T (suggesting a value of around 21\,T
at 4.2\,K) were reported. The higher anisotropy of the undoped
tape is expected from its higher $T_\mathrm{c}$ (34.4\,K) and also
$B_\mathrm{c2}$ tends to decrease for transition temperatures of
below $\sim 32$\,K \cite{Eis07rev}. The larger $I_\mathrm{c0}$ of
the undoped tape confirms that carbon doping decreases pinning
\cite{Eis07rev,Kim08} and the connectivity.

A discrepancy between experimental and calculated data was found
at small currents ($< 1$\,A, $\theta_\mathrm{t}=0^\circ$, not
shown in Fig.~\ref{FigJcexp}), which might be a consequence of the
finite voltage during transport measurements (1\,$\mu$Vcm$^{-1}$),
which allows small currents passing normal conducting grains. Such
normal currents are excluded in the present model and might add a
significant contribution at small critical currents (by connecting
otherwise disconnected superconducting clusters), but they can
certainly be neglected at high critical currents. On the other
hand, the discrepancy could also result from inhomogeneities,
texture gradients \cite{Son02,Lez06b}, or a different distribution
function of the grain orientation. Note that the Gaussian
distribution function was also used in Ref.~\onlinecite{Lez06b}
and an even higher degree of texture ($\alpha_\mathrm{t}\sim
20^\circ$ and $\sim 25^\circ$ according to our definition) was
found near the sheath of cold rolled ex-situ tapes, decreasing
towards the center.

The angular dependence of $J_\mathrm{c}$ was calculated with the
same parameters, which describe the data very well. However,
without a knowledge of the anisotropy the texture
($\alpha_\mathrm{t}$) cannot be derived due to the mutual
influence of $\alpha_\mathrm{t}$ and $\gamma$.

\section{Conclusions}

The proposed model successfully describes the field and angular
dependence of the critical currents in weakly textured MgB$_2$
tapes and explains the large difference between the
\emph{intrinsic} and \emph{extrinsic} ($J_\mathrm{c}$) anisotropy.

\begin{acknowledgments}
The authors wish to thank Professor Harald Weber for critically
reading the manuscript and for his valuable comments. We are
grateful to Asger Abrahamsen for his synchrotron measurements and
fruitful discussions.
\end{acknowledgments}


\end{document}